\documentclass[
    aps,prd,reprint,            % PRX format 
    superscriptaddress,         % use superscript for addresses
    amsfonts,amsmath,amssymb,   % load AMS packages
]{revtex4-2}

\usepackage{braket} 
\usepackage{graphicx}
\usepackage{dcolumn}
\usepackage{multirow}
\usepackage{verbatim}
\usepackage{natbib}
\usepackage{graphicx}
\usepackage{amsmath,amssymb}
\usepackage{bm}
\usepackage{dsfont}
\usepackage{cancel}
\usepackage{float}
\usepackage{braket}
\usepackage{siunitx}
\usepackage{xspace}
\usepackage{comment}
\usepackage{makecell}
\usepackage{xcolor}
\usepackage[normalem]{ulem}

\usepackage[colorlinks=true,citecolor=blue,urlcolor=blue,linktocpage=true,linkcolor=blue]{hyperref}
\usepackage[capitalize]{cleveref}
\usepackage{orcidlink}
\usepackage[absolute,overlay]{textpos}

\newcommand*{\DM}{\ensuremath{\mathrm{DM}}}
\newcommand*{\DR}{\ensuremath{\mathrm{DR}}}
\newcommand*{\nueff}{\ensuremath{\nu_{\mathrm{eff}}}}

\DeclareSIUnit \parsec {pc}

\begin{document}
\begin{textblock*}{\textwidth}[1,0](200mm,10mm)
\noindent
\flushright
\text{MITP-25-030}
\end{textblock*}

\preprint{APS/123-QED}

\title{Generalized neutrino isocurvature}

\author{Christopher Gerlach\,\orcidlink{0009-0000-3431-2174}}
\email{cgerlach@uni-mainz.de}
\affiliation{%
 PRISMA$^{+}$ Cluster of Excellence \& Mainz Institute for Theoretical Physics,\\
 Johannes Gutenberg-Universit\"at Mainz, 55099 Mainz, Germany
}%

\author{Wolfram Ratzinger\,\orcidlink{0000-0003-2312-7154}}
\email{wolfram.ratzinger@weizmann.ac.il}
\affiliation{Department of Particle Physics and Astrophysics,\\
Weizmann Institute of Science, Rehovot, Israel 7610001}

\author{Pedro Schwaller\,\orcidlink{0000-0002-6307-3069}}%
\email{pedro.schwaller@uni-mainz.de}
\affiliation{%
 PRISMA$^{+}$ Cluster of Excellence \& Mainz Institute for Theoretical Physics,\\
 Johannes Gutenberg-Universit\"at Mainz, 55099 Mainz, Germany
}%

\date{\today}

\begin{abstract}

Searches for neutrino isocurvature usually constrain a specific linear combination of isocurvature perturbations. In this work, we discuss realistic cosmological scenarios giving rise to neutrino isocurvature.
We show that in general both, neutrino and matter isocurvature perturbations are generated, whose ratio we parameterize by a newly introduced mixing angle. We obtain the first limits on this new mixing angle from PLANCK data, and discuss novel insights into the early Universe that could be provided by future measurements. 
\end{abstract}

\maketitle

%%%%%%%%%%%%%%%%%%%%%%%%%%%%%%%%%%%%%%%%%%%%%%%%%%%%%%%%%%%
\section{Introduction}
%%%%%%%%%%%%%%%%%%%%%%%%%%%%%%%%%%%%%%%%%%%%%%%%%%%%%%%%%%%

At the emission of the cosmic microwave background (CMB), four fluids contribute to the energy of the Universe: photons, baryons, neutrinos and dark matter. 
Taking a bottom-up perspective, it is therefore natural to assume that there are four independent initial conditions for perturbations to the homogeneous and isotropic background cosmology. 
In terms of gauge invariant quantities, they may for example be given by a curvature perturbation $\zeta$ and isocurvature perturbations between the photons and the other fluids \cite{Bardeen:1980kt}, 
\begin{align}
    S_{\gamma X}=3(\zeta_\gamma-\zeta_X)=-3H\left( \frac{\delta\rho_\gamma}{\dot\rho_\gamma}-\frac{\delta\rho_X}{\dot\rho_X}\right)\,,
\end{align}
where $X\in\{\nu,\DM,\mathrm{b}\}$, $H$ is the Hubble parameter and $\delta\rho_{\gamma,X}$ the energy density perturbation of the background energy density $\rho_{\gamma,X}$.

To this day, all observations of anisotropies are in agreement with the existence of only a non-vanishing initial curvature perturbation $\zeta\neq0\,, S_{\gamma X}=0$\,. 
In terms of cosmological histories this is easily motivated. 
In any scenario where there is only one single degree of freedom during inflation or at some point after inflation, i.e., the whole energy content thermalizes, this so called adiabatic initial condition is realized \cite{Weinberg:2004kf,Weinberg:2004kr}. 

\begin{figure}[!b]
\centering
\includegraphics[width=0.45\textwidth]{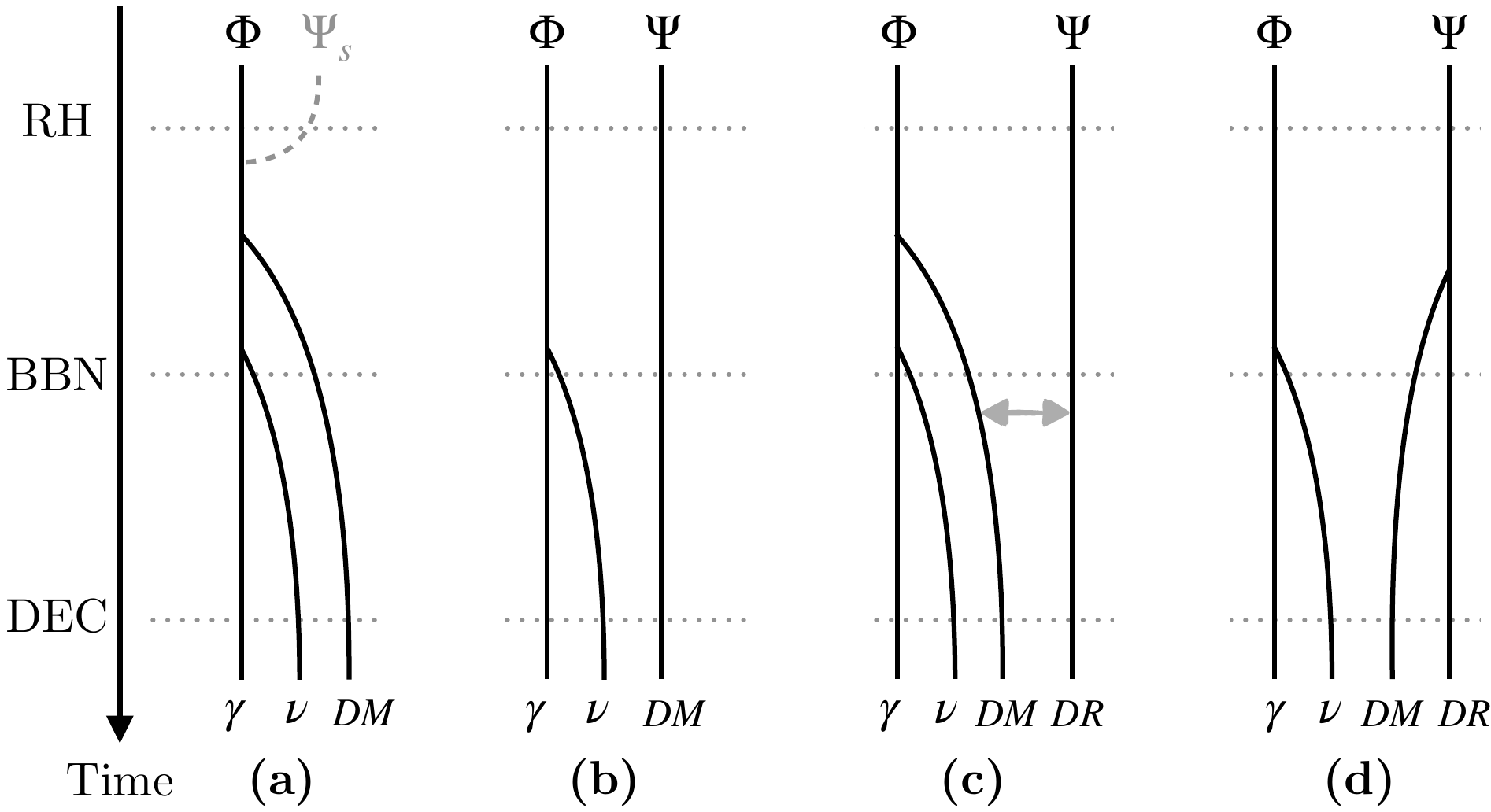}
\caption{Cosmic histories in comparison: The lines correspond to energy flow in the corresponding sector or fluid. Scenario (a) is the standard adiabatic case. Scenario (b) may lead to dark matter isocurvature but not to neutrino isocurvature (both compared to photons). In (c) a second sector adds a dark radiation background, which may entail neutrino isocurvature. The arrow indicates the possibility of an DM-DR interaction. In (d), a dark sector $\Psi$ sources not only dark radiation, but also dark matter.}
\label{fig:cosmological_histories}
\end{figure}

From a top-down perspective, constraining isocurvature therefore corresponds to testing these most simple cosmologies, while the discovery of isocurvature would point to a much richer cosmology, since throughout its evolution there must be multiple non-thermalized sectors present. 
Given that so far only curvature perturbations have been observed, it seems plausible that there is a hierarchy between the different perturbations. 
If this is the case, going from a cosmological history where at early times only one degree of freedom was of relevance, as a first step, it seems logical to consider one where there are two independent sectors at all times leading to one combination of isocurvature fluctuations besides the adiabatic one.
The immediate questions that come up in this context are which cosmological history leads to which linear combination of $S_{\gamma X}$ and whether any of these combinations are particularly well motivated. In an attempt to answer this question, we consider cosmological histories of increasing complexity in \cref{sec:overview neutrino isocurvature} (cf. \cref{fig:cosmological_histories}). We find that cosmologies with neutrino isocurvature generically feature fully correlated admixtures of matter isocurvature.

From the bottom-up instead, one may try to keep a search for the four independent initial conditions mentioned above as general as possible.
This corresponds to generalizing the curvature power spectrum $\mathcal{P}_{\zeta}(k)$ which quantizes the correlator $\langle \zeta(k)\zeta(k)\rangle$ into a symmetric 4x4 matrix that further gives the correlators $\langle \zeta(k) S_{\gamma X}(k)\rangle$ and $\langle  S_{\gamma X}(k)S_{\gamma Y}(k)\rangle$\,\cite{Bucher:1999re}.
While $\mathcal{P}_{\zeta}(k)$ is commonly given in terms of a normalization and a tilt, so two free parameters, doing the same with all other entries of this symmetric matrix would result in 18 additional free parameters \footnote{The correlation matrix has to be kept positive definite. See Ref.~\cite{Planck:2013jfk} for details on how to ensure this constraint.}.
While for WMAP, this most general search has been carried out \cite{Bucher:2000hy,Bucher:2004an,Moodley:2004nz,Bean:2006qz}, this is not the case for the best data available, currently coming from Planck \cite{Planck:2018jri}.

The Planck analysis \cite{Planck:2013jfk,Planck:2015sxf,Planck:2018jri} has instead considered generalizations where besides $\zeta\neq0$, one additional linear combination of the $S_{\gamma X}$ is non-zero. 
With this restriction, the power spectrum is generalized to a 2x2 matrix and with the same counting as above one introduces four additional parameters only. To such a search our work on the origin of particular combinations of isocurvature is of immediate importance, since it allows one to interpret the results in terms of specific cosmological histories.
The insight from \cref{sec:overview neutrino isocurvature} that neutrino isocurvature is generically accompanied by arbitrary amounts of matter isocurvature, motivates us to conduct a search for this generalized isocurvature using CMB and large scale structure (LSS) data, where $S_{\gamma\nu}/S_{\gamma \DM}$ is left as a free parameter, in \cref{sec:CMB fit}\,.

%%%%%%%%%%%%%%%%%%%%%%%%%%%%%%%%%%%%%%%%%%%%%%%%%%%%%%%%%%%
\section{Overview on cosmic histories}
\label{sec:overview neutrino isocurvature}
%%%%%%%%%%%%%%%%%%%%%%%%%%%%%%%%%%%%%%%%%%%%%%%%%%%%%%%%%%%

To see whether
any specific combinations of isocurvature modes
are particularly well motivated, we compare cosmologies of increasing complexity in \cref{fig:cosmological_histories}.
The first one, (a), is the standard single-field inflation scenario with inflaton $\Phi\,$. In this model, at early times there can only be a time difference between the equal energy slicing of the inflaton sector with respect to spatially flat slices, which leads to the fluctuations being adiabatic. The fluctuations then stay adiabatic throughout all of the following evolution. This holds true for much more complicated inflationary sectors as long as they reheat a common fluid at some point \cite{Weinberg:2004kr,Weinberg:2004kf}. An example might be multiple fields $\Psi_s$ driving inflation as illustrated in \cref{fig:cosmological_histories}.

The following three examples all feature two independent, non-thermalized fluids throughout the evolution from inflation until today and therefore feature besides curvature fluctuations one independent linear combination of isocurvature. In example (b), the second sector $\Psi$ later on acts as cold dark matter (DM). A minimal example would be scalar DM produced via the misalignment mechanism \cite{Preskill:1982cy,Abbott:1982af,Dine:1982ah,Linde:1984ti,Hamann:2009yf,Caputo:2023ikd}. Since the photons and neutrinos are in thermal equilibrium at the time when they decouple, the relative isocurvature between them is $S_{\gamma\nu}=0\,$. One therefore only has DM isocurvature $S_{\gamma \DM}\neq 0\,$, as one may have naively expected.

In (c) and (d), we show examples of cosmologies that realize a non-zero neutrino isocurvature. As just discussed, neutrinos are in thermal equilibrium at times well before big bang nucleosynthesis (BBN). 
In order to achieve fluctuations in the neutrino density one must therefore either consider fluctuations in the lepton number or consider a separate sector that at times after BBN contributes to the neutrinos. 
We will discuss both possibilities in the following.

%%%%%%%%%%%%%%%%%%%%%%%%%%%%%%%%%%%%%%%%%%%%%%%%%%%%%%%%%%%
\subsection{Neutrino isocurvature from a dark sector}
%%%%%%%%%%%%%%%%%%%%%%%%%%%%%%%%%%%%%%%%%%%%%%%%%%%%%%%%%%%

The simplest extension to produce neutrino isocurvature is the setup presented in \cref{fig:cosmological_histories} (c). A second scalar field $\Psi$ next to the inflaton, which therefore can have uncorrelated quantum perturbations before inflation, later decays to dark radiation which at the time of CMB emission is free-streaming and does not interact with other fluids. This dark sector would necessarily carry uncorrelated isocurvature perturbations.

While neutrino masses are expected to become relevant to LSS surveys in the near future \cite{Lesgourgues:2006nd,Lattanzi:2017ubx,DeSalas:2018rby}, to this day, neutrinos may be modeled as free-streaming radiation, only interacting gravitationally at all times after BBN. 
Neutrinos and the dark sector are then indistinguishable to CMB and LSS probes. To this end, we define an effective neutrino density and isocurvature perturbation 
\begin{align}
    \rho_{\nueff}=\rho_{\nu}+\rho_{\rm DR}\,,\quad S_{\gamma\nueff}=\frac{\rho_{\rm DR}}{\rho_{\nueff}}S_{\gamma\DR}\,,
\end{align}
where we assumed $S_{\gamma\nu}=0$.
One may naively think that scenario (c) in the absence of direct interaction (indicated by gray arrows in \cref{fig:cosmological_histories}) corresponds to ${S}_{\gamma\nueff}\neq0\,$ and ${S}_{\gamma \DM} = 0$, while (d) corresponds to ${S}_{\DM\,\mathrm{DR}}=0$, which is equivalent to $S_{\gamma\DM}=\rho_{\nueff}/\rho_\DR\,S_{\gamma \nueff}$.

It is obvious that if we allow for direct interactions between the fluids, any ratio of $S_{\gamma\nueff}/S_{\gamma\DM}$ may be achieved. For example if we introduce in (c) an interaction that partially thermalizes the DM and DR sector, $S_{\DM\,\DR}$ would be reduced, and depending on the interaction strength, one can interpolate to any value of $S_{\gamma\nueff}/S_{\gamma\DM}$ between the ones of example (c) and (d). 

In an upcoming paper \cite{Gerlach:2025uxo}, we will further concentrate on example (c) and show that, contrary to the naive expectation, even in the absence of direct interactions, a non-zero $S_{\gamma\DM},\,S_{\gamma\mathrm{b}}$ can be induced. Using the separate universe approach \cite{Wands:2000dp,Lyth_2003,Lyth_2005,Artigas_2022} one can roughly explain the effect as follows: In the presence of dark radiation isocurvature on super horizon scales, the Universe may be viewed as a collection of patches with varying dark radiation density at a given density in the standard model sector (the one originating from $\Phi$ in \cref{fig:cosmological_histories} (c)). In most models of dark matter production \cite{Cirelli:2024ssz} or baryogenesis \cite{Dine:2003ax}, the yield crucially depends on the ratio between particle physics rates and the Hubble rate. The particle physics rates themselves only depend on the density in the standard model sector (which are usually both given as a function of temperature) and are therefore the same for all patches with the above slicing. The Hubble rate, on the other hand, depends on the energy density in both sectors and therefore varies. This results in varying yields in different patches, a non-vanishing matter isocurvature. Similar effects have been discussed in the context of the curvaton in Refs.~\cite{Lyth:2002my,Lyth:2003ip}, and Ref.~\cite{Adshead:2020htj} pointed out that the dark radiation scenario leads to varying element abundances during BBN in the same way.

It should be noted that in the purely gravitationally coupled cases, the amount of induced matter isocurvature is proportional to the relative energy density in the dark sector $\Omega_{\DR}$ at the time of dark matter production and baryogenesis, respectively. The effect might therefore be suppressed if e.g. the dark sector behaves matter-like initially and only later on acts as radiation, resulting in $S_{\gamma \DM}\simeq0$. This is also the reason why these complications do not appear in case (b) of dark matter isocurvature. Let us also mention that the amount of dark radiation is bounded at the 95\% confidence level $\rho_\DR/\rho_\nu\lesssim 0.1$ from CMB observations \cite{Planck:2018vyg}. Below we present our bounds in terms of $S_{\gamma \nueff}$ such that small dark radiation densities may be equivalently compensated by larger fluctuations $S_{\gamma \DR}$\,.

We conclude that besides this specific scenario resulting in $S_{\gamma \DM}=0$, all other cases featuring neutrino isocurvature lead to some range of $S_{\gamma \nu}/S_{\gamma \DM}$ and therefore motivate a generalized search where this ratio is scanned over.

%%%%%%%%%%%%%%%%%%%%%%%%%%%%%%%%%%%%%%%%%%%%%%%%%%%%%%%%%%%%%%%%%%%%
\subsection{Neutrino isocurvature from lepton number fluctuations}
%%%%%%%%%%%%%%%%%%%%%%%%%%%%%%%%%%%%%%%%%%%%%%%%%%%%%%%%%%%%%%%%%%%%
Let us briefly comment on the scenario without dark radiation present at CMB emission.
A new physics sector which decays into highly relativistic neutrinos after they decouple at around $T\approx \SI{1}{MeV}$ can be treated in analogy to the previous discussion by identifying the energy in neutrinos resulting from the decay with the free-streaming dark radiation from before.

A lepton number conserving decay at earlier times only leads to an increase of the temperature in the primordial plasma and would not induce neutrino isocurvature.
If the decay induces fluctuations in the lepton number to photon ratio, this may still lead to fluctuations in the neutrino energy density.
Since this dependence starts at quadratic order only, a sizable background lepton asymmetry is required in order for its fluctuations to induce perturbations in the energy density at linear order \cite{Lyth:2002my,DiValentino:2011sv}. 
The fluctuations in the energy density again lead to the gravitational effects that induce matter fluctuations. 
Furthermore, if the large lepton number is linked to the generation of the observed baryon asymmetry, one would expect an imprint of fluctuations on the baryon number as well. In scenarios where the SM plasma reaches temperatures well above the electroweak scale, sphaleron processes equilibrate the lepton and baryon asymmetry, favoring a tiny lepton asymmetry \cite{Harvey:1990qw}. The generation of the baryon asymmetry would need to be a much weaker residual effect.
We therefore reach the same conclusion that neutrino isocurvature is in general accompanied by matter isocurvature. Note that the lepton asymmetry is bounded by CMB and BBN observations~\cite{DiValentino:2011sv,Domcke:2025jiy} paralleling the bound on $\rho_\DR$.

%%%%%%%%%%%%%%%%%%%%%%%%%%%%%%%%%%%%%%%%%%%%%%%%%%%%%%%%%%%
\subsection{Basis choice for mixed isocurvature}
%%%%%%%%%%%%%%%%%%%%%%%%%%%%%%%%%%%%%%%%%%%%%%%%%%%%%%%%%%%
Let us now return to the question whether any specific linear combinations of isocurvature are particularly well motivated. 
First note that the so called compensated matter perturbation, which is defined by the 
isocurvature of the total matter content with respect to the total radiation vanishing and corresponds to
\begin{equation}
    \zeta=0\,,\quad S_{\gamma\nu}=0\,,\quad S_{\gamma\DM}=-\frac{\rho_b}{\rho_\DM}S_{\gamma b}\,,
\end{equation}
is not observable at linear order in the CMB \cite{Grin:2011tf,Grin:2011nk}. It is therefore 
only constrained at amplitudes much larger than the observed adiabatic ones \cite{Planck:2018jri,Lee:2021bmn,Barreira:2023uvp}. In this paper we restrict ourselves to magnitudes of isocurvature smaller than the observed CMB fluctuations, $S\lesssim 10^{-5}$. This allows us to recast any fluctuation of baryons and dark matter into an effectively unconstrained compensated matter perturbation and an overall matter fluctuation that may be expressed by a dark matter fluctuation, while choosing $S_{\gamma b}=0$\,. This justifies our omission of baryon isocurvature for the most part in the discussion so far and reduces the question of what isocurvature perturbation to consider to a choice of the ratio $S_{\gamma\nu}/S_{\gamma \DM}$\,.

In general, this scenario can therefore be parametrized by the isocurvature fraction relative to the adiabatic fluctuation,
\begin{equation}
    f_{\rm iso}^2=\frac{\langle S_{\gamma\nu}^2\rangle +\langle S_{\gamma\DM}^2\rangle}{\langle\zeta_{\rm tot}^2\rangle}\,,
\end{equation}
the ratio between the fully correlated neutrino and matter isocurvature that we will give in terms of an angle
\begin{equation}
    \tan(\varphi)=\frac{S_{\gamma\nu}}{S_{\gamma\DM}}\,,
\end{equation}
as well as the possible correlation between the isocurvature and the adiabatic mode that we may take with respect to the dark matter component,
\begin{equation}
    c_{\zeta S}=\frac{\langle S_{\gamma\DM} \zeta_{\rm tot}\rangle}{\sqrt{\langle S_{\gamma\DM}^2\rangle\langle\zeta_{\rm tot}^2\rangle}}\,.
\end{equation}

Previous works \cite{Kawasaki:2011rc,DiValentino:2011sv,Kawakami:2012ke,Planck:2018jri,Adshead:2020htj,Ghosh:2021axu} have only considered specific choices of the angle $\varphi$\,.
The initial condition called dark matter isocurvature corresponds to $\varphi=0$ with our definitions.  
The one referred to as neutrino isocurvature in the literature is however arranged such that there is no isocurvature between the total radiation and matter. 
This corresponds to the combination $\varphi~=~\arctan[(1+\rho_{\gamma}/\rho_{\nu})/(1+\rho_b/\rho_\DM)]\approx1.12$\,. Further the normalization is chosen such that the isocurvature fraction $f_\mathrm{iso}$ needs to be multiplied by a factor of $1.4$ to
compare to our convention. We will refer to this choice of $\varphi$ as compensated neutrino isocurvature in analogy to the compensated matter perturbation. Conversely, we will refer to pure neutrino isocurvature as $S_{\gamma\DM}=0,\,\varphi=\pi/2$\,, which results from the scenario shown in \cref{fig:cosmological_histories}~(c) if direct and gravitational interactions between the sectors are negligible.

%%%%%%%%%%%%%%%%%%%%%%%%%%%%%%%%%%%%%%%%%%%%%%%%%%%%%%%%%%%
\section{Constraining cosmologies from isocurvature}
\label{sec:CMB fit}
%%%%%%%%%%%%%%%%%%%%%%%%%%%%%%%%%%%%%%%%%%%%%%%%%%%%%%%%%%%
From the above discussion, it should be clear that in cosmologies that feature neutrino isocurvature $S_{\gamma\nu}\neq 0$, generically non-vanishing and fully correlated matter fluctuations can be expected. 
Below we derive the imprint of a scenario with arbitrary $\varphi$ on the CMB and LSS in order to carry out a search using existing data.

%%%%%%%%%%%%%%%%%%%%%%%%%%%%%%%%%%%%%%%%%%%%%%%%%%%%%%%%
\subsection{Computation of cosmological observables}
%%%%%%%%%%%%%%%%%%%%%%%%%%%%%%%%%%%%%%%%%%%%%%%%%%%%%%%%
In order to find the effect of isocurvature on the CMB and other cosmological observables, one needs to solve the evolution equations of the perturbations as the relevant scales enter the horizon. This is commonly done in a fixed gauge. The isocurvature mode in question then corresponds to a set of initial conditions that one has to determine in the gauge of choice.
Further, since the evolution equations typically feature singularities as $\tau\rightarrow0$, one has to analytically expand the solution to finite conformal time $\tau$ before the equations can be solved numerically. In \cref{sec:initial conditions} we give this expansion in synchronous gauge for $S_{\gamma\DM}=1,\,S_{\gamma\nu}=0$ and $S_{\gamma\DM}=0,\,S_{\gamma\nu}=1$\,. The general case may then be obtained by adding the solutions with weights $\cos(\varphi)$ and $\sin(\varphi)$\,, respectively. This approach is straightforward to generalize to more exotic scenarios containing e.g. self-interacting instead of free-streaming dark radiation \cite{Ghosh:2021axu}.

\begin{figure}
    \centering
    \includegraphics[width=\columnwidth]{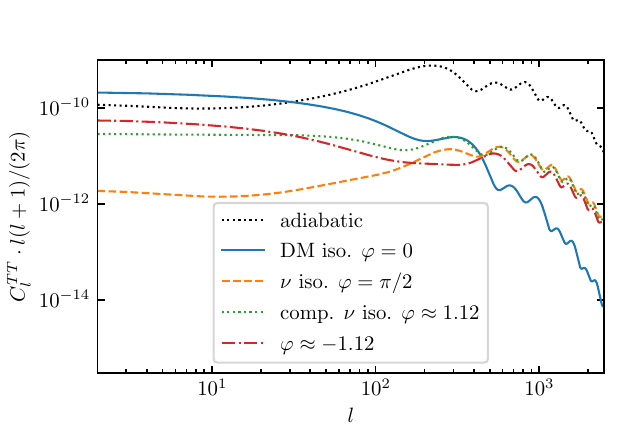}
    \caption{Comparison of temperature correlations in the CMB as a function of angular multipole $l$ resulting from adiabatic, dark matter and neutrino isocurvature only as well as two admixed cases. For simplicity, we take all primordial power spectra to be flat and with the same amplitude.}
    \label{fig:TTspectra}
\end{figure}

We use a modified version of the code CLASS \cite{Blas:2011rf} to solve the evolution equations of the perturbations and compute all relevant CMB and LSS observables. In \cref{fig:TTspectra} we show the spectrum of lensed temperature correlations in the CMB for a fiducial $\mathrm{\Lambda CDM}$ cosmology using the best fit parameters from Planck \cite{Planck:2018vyg}, but for simplicity taking a flat primordial spectrum, i.e., setting the scalar index $n=1$\,. Besides the adiabatic initial conditions that are dominating our Universe, we show four different scenarios with isocurvature instead of the adiabatic mode at the same amplitude. 

It becomes apparent that the resulting spectra differ in the position of the acoustic peaks as well as their heights. It is therefore clear that CMB data is able to distinguish between different values of $\varphi$\,. In the event of detection, $\varphi$ can therefore be determined and would serve as an invaluable piece of information when trying to disentangle the various cosmologies discussed above. On the other hand, since the signatures differ quiet drastically, it is easy to imagine that an isocurvature signal may be missed when narrowing the search to one $\varphi$ only. The same conclusion is reached in \cref{sec:CMB polarization and LSS}, where we consider CMB polarization and LSS.

%%%%%%%%%%%%%%%%%%%%%%%%%%%%%%%%%%%%%%%%%%%%%%%%%%%%%%%%
\subsection{Fit to CMB and LSS data}
%%%%%%%%%%%%%%%%%%%%%%%%%%%%%%%%%%%%%%%%%%%%%%%%%%%%%%%%
\begin{figure}
    \centering
    \includegraphics[width=\columnwidth]{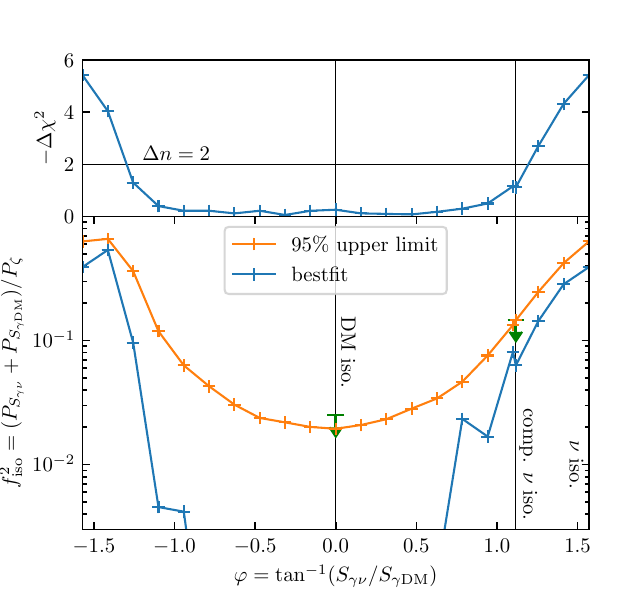}
    \caption{Upper bound on the isocurvature fraction as a function of the mixing angle $\varphi$ between neutrino and matter isocurvature (orange line). The green arrows represent the bounds found in Ref.~\cite{Planck:2018jri}. For mixing angles close to $\varphi=\pi/2\sim-\pi/2$, corresponding to vanishing matter isocurvature, the fit prefers non-zero fractions (blue line) leading to a reduction in $\chi^2$ (top panel).}
    \label{fig:upperBound}
\end{figure}

We carry out the first search for generalized neutrino isocurvature using Planck 2018 (including TTTEEE + lensing \cite{Planck:2019nip}) as well as baryon acoustic oscillation data (including BOSS DR12 \cite{BOSS:2016wmc} + MGS \cite{Ross:2014qpa} + 6dFGS \cite{Beutler:2011hx}). 
We perform a Markov chain Monte Carlo (MCMC) analysis using MontePython \cite{Audren:2012wb,Brinckmann:2018cvx}. 
Additionally to the base-$\Lambda$CDM parameters from Ref.~\cite{Planck:2018vyg}, we fit the effective number of neutrinos as all models discussed in \cref{sec:overview neutrino isocurvature} involve additional degrees of freedom. 
We here present the results of a series of runs in which the spectral index $n_\mathrm{iso}=1$ is fixed and $\varphi$ is varied from $-\pi/2$ to $\pi/2$ in between the fits. 
In each run, $f_\mathrm{iso}$ and $c_{\zeta S}$ are fitted.
We use a linear-flat prior for $f_\mathrm{iso}$ preferring larger values than a log-flat prior.
Our results therefore can be interpreted as conservative bounds on the isocurvature mode from inflationary scenarios which lead to $n_\mathrm{iso}\approx 1$\,
\footnote{We made some initial runs in which the spectral index $n_\mathrm{iso}$ was varied freely as well. Due to the absence of significant evidence for isocurvature these runs did not converge. Therefore we only present bounds on the simpler yet well motivated scenario $n_\mathrm{iso}= 1$.}.

In \cref{fig:upperBound} we show the dependence of the 95 percentile of $f_\mathrm{iso}^2=(P_{S_{\gamma\nu}}+P_{S_{\gamma\DM}})/P_{\zeta}|_{k_\mathrm{piv}}$ as a function of $\varphi$\, in orange. For comparison the bounds found in Ref.~\cite{Planck:2018jri} are indicated by green arrows. As the analysis of Ref.~\cite{Planck:2018jri} allowed the isocurvature spectral index to vary, we compare our results to the most stringent one out of the three scales considered in Tab.~14 of Ref.~\cite{Planck:2018jri}, the one at $k_\mathrm{low}\,$. More specifically, we took the values $f_\mathrm{iso}^2=2.5\%$ and $7.4\%$ for the dark matter and compensated neutrino case, respectively, where for the latter one we accounted for the difference in normalization by multiplying with $(1.4)^2$. Our results agree with Ref.~\cite{Planck:2018jri} to within 30\%, where the discrepancies are most likely due to differences in analysis.

We find that the bound on the isocurvature fraction is weaker for scenarios with vanishing matter isocurvature $\varphi\approx\pi/2\sim-\pi/2$\,. This may be contributed to the fact that the signatures in the case of $S_{\gamma \DM}=0$ are suppressed on all scales relative to the adiabatic ones as can be seen from \cref{fig:TTspectra} (see also \cref{sec:CMB polarization and LSS}).  Furthermore, we find that the fit shows a slight preference for non-vanishing isocurvature for these values of $\varphi$ as the reduction of $\chi^2$ compared to a fit of a $\Lambda$CDM cosmology with varying $N_\mathrm{eff}$ exceeds the number $\Delta n=2$ of additional fit parameters ($f_\mathrm{iso}$, $c_{\zeta S}$).

\begin{figure}[h]
    \centering
    \includegraphics[width=\columnwidth]{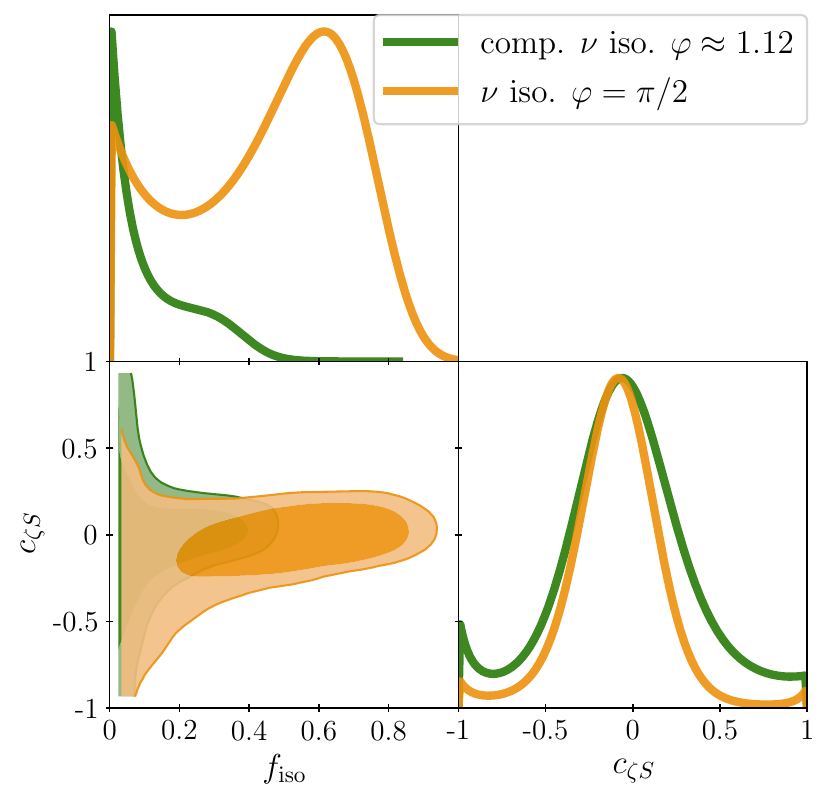}
    \caption{Inferred values from the fit of the isocurvature fraction and correlation with the adiabatic mode in the case of neutrino isocurvature ($S_{\gamma\DM}=0$, orange) and compensated neutrino isocurvature (green).}
    \label{fig:triangle}
\end{figure}

In \cref{fig:triangle} we compare the posterior distribution as found by the MCMC projected onto the plane of isocurvature fraction $f_\mathrm{iso}$ and correlation with the adiabatic mode $c_{\zeta S}$ for compensated ($\varphi\approx 1.12$) and pure ($\varphi=\pi/2$) neutrino isocurvature. 
We find that in the case of no correlation and pure neutrino isocurvature, a non-zero isocurvature fraction is preferred, while for the compensated neutrino isocurvature used by the Planck collaboration, this effect is not visible, emphasizing again the importance of mixed isocurvature modes.

Further we note that correlation with the adiabatic mode is disfavored in both cases, in agreement with  Ref.~\cite{Planck:2018jri}.
In the case of curvaton models predicting fully correlated ($c_{\zeta S}=\pm1$) isocurvature \cite{Lyth:2002my,Lyth:2003ip,Kawasaki:2011rc,DiValentino:2011sv}, the bounds on the isocurvature fraction are therefore significantly more stringent.

Again we want to stress that, if this hint could be substantiated with additional data, only a small range of values $\varphi$ would lead to the observed signatures. Such a measurement of $\varphi$ would serve as an important puzzle piece when trying to decipher our cosmos' origins.

\begin{acknowledgments}
We would like to thank J. Lesgourgues for insights and discussions on MontePython.
The authors gratefully acknowledge the computing time granted on the supercomputer MOGON 2 at Johannes Gutenberg University Mainz (hpc.uni-mainz.de). PS and CG acknowledge support by the Cluster of Excellence “Precision Physics, Fundamental Interactions, and Structure of Matter” (PRISMA+ EXC 2118/1 and PRISMA++ EXC 2118/2) funded by the Deutsche Forschungsgemeinschaft (DFG, German Research Foundation) within the German Excellence Strategy (Project No. 390831469).
\end{acknowledgments}

\appendix

%%%%%%%%%%%%%%%%%%%%%%%%%%%%%%%%%%%%%%%%%%%%%%%%%%%%
\section{Initial conditions in coordinates}
\label{sec:initial conditions}
%%%%%%%%%%%%%%%%%%%%%%%%%%%%%%%%%%%%%%%%%%%%%%%%%%%%

In this appendix, the 
initial conditions are provided for the different isocurvature setups referred to above. Isocurvature and curvature perturbations translate to metric and energy contrast perturbations at horizon entry. 
These initial conditions, combined with the perturbed Einstein-Boltzmann equations for the given species, can be solved to follow the subsequent subhorizon evolution of the perturbations. The solutions for the setups are provided in synchronous gauge and we follow the notation from Ref.~\cite{Bucher:1999re}.
They can be forwarded to the Boltzmann code of choice.

In the adiabatic scenario, the only non-vanishing initial value is the metric perturbation $\eta(\tau=0)=1$. Note that this result is subject to the normalization $\zeta=-1$.

In order for the metric perturbations not to diverge as $\tau\rightarrow 0$ one has to ensure that the radiation fluid is homogeneous during radiation domination \cite{Bucher:1999re,Ghosh:2021axu}.
One therefore sets
\begin{align}
    \delta\rho_\gamma(0)+\delta\rho_{\nueff}(0)=0\,.
    \label{eq:gauge_condition_radiation}
\end{align}
This equation together with a set of
\begin{align}
    S_{\gamma X}=\frac{3}{4} \delta_\gamma(0)-\alpha_X\delta_X(0)\,,
\end{align}
where $X\in\{\nueff,\DM,\mathrm{b}\}$ and $\alpha_{\nueff}=3/4,\,\alpha_{\DM,\mathrm{b}}=1$ then fully specifies the initial condition.

For pure dark matter isocurvature, corresponding to $\varphi=0$,
the gauge-invariant initial conditions are given by
\begin{align}
    &S_{\gamma\DM}=\mathcal{N}\,,\\
    &S_{\gamma\nueff}=0\,, \\
    &S_{\gamma \mathrm{b}}=0\,.
\end{align}
This is fulfilled by $\delta_\DM(0)=1$ with the other $\delta_i(0)=0$ and corresponds to the normalization $\mathcal{N}=-1$. This is the cold dark matter isocurvature scenario from Ref.~\cite{Bucher:1999re}. 

Pure neutrino density isocurvature includes neutrinos as well as free-streaming dark radiation. It is represented by the mixing angle $\varphi=\pi/2$. The gauge-invariant initial conditions here read
\begin{align}
    &S_{\gamma\nueff}=\mathcal{N}\,,\\
    &S_{\gamma\DM}=0\,, \\
    &S_{\gamma \mathrm{b}}=0\,.
\end{align}

The gauge-independent ansatz then translates into
\begin{align}
    & \delta_{\nueff}(0)=\frac{4}{3}\frac{\rho_{\gamma}}{\rho_{\mathrm{rad}}}\,, \\
    & \delta_\gamma(0)=-\frac{4}{3}\frac{\rho_{\nueff}}{\rho_{\mathrm{rad}}}\,,  \\
    & \delta_\DM(0)=\delta_\mathrm{b}(0) =\frac{3}{4}\delta_\gamma(0)\,.
\end{align}
This approach makes sure that $S_{\gamma\DM}=S_{\gamma\mathrm{b}}= 0$ and only the neutrino isocurvature does not vanish.

\begin{table*}[]
    \centering
    \begin{tabular}{|c|c|c|c|c|c}
        \hline
        Pert. & $\mathcal{O}(1)$ & $\mathcal{O}(\tau)$ & $\mathcal{O}(\tau^2)$ & $\mathcal{O}(\tau^3)$ \\
        \hline \hline
        $h$ && $ -R_\nu \omega $ & $\frac{3}{8} R_\nu \omega^2 $ & 
        $ \frac{R_\nu k^2\omega }{90} (3 R_\mathrm{b}+4) -\frac{R_\nu \omega^3}{8} $ 
        \\
        \hline
        $\eta$ &  & $\frac{R_\nu \omega}{6}$ & $\frac{2
   (R_\nu-1) R_\nu k^2}{9 (4 R_\nu+15)}-\frac{R_\nu
   \omega^2}{16} $ & 
   \begin{tabular}{@{}c@{}} $\frac{R_\nu
   \omega^3}{48} -\frac{R_\nu 2 R_\mathrm{b}  k^2 \omega}{360} $ \\ $ -\frac{R_\nu (16 R_\nu (3 R_\nu+40)-175) k^2 \omega}{360 (2 R_\nu+15) (4
   R_\nu+15)}$  \end{tabular} \\
        \hline
        $\delta_{\gamma}$ & $-\frac{4 R_\nu}{3}$  & $\frac{2 R_\nu \omega \tau
   }{3}$  & $\frac{1}{36} R_\nu \left(8 k^2-9 \omega^2\right)$  &  \begin{tabular}{@{}c@{}} $ \frac{R_\nu \omega^3}{12}-\frac{6 R_\nu  k^2 \omega}{90} $ \\ $ -\frac{2 R_\nu (R_\mathrm{b}
   (R_\nu-6)) k^2 \omega}{90
   (R_\nu-1)}$  \end{tabular} \\
         \hline
        $\theta_{\gamma \mathrm{b}}/k$ && $-\frac{R_\nu k}{3} $ & $ -\frac{R_\nu R_\mathrm{b} k\omega}{4(R_\nu-1) + \frac{R_\nu k\omega}{12}}$ &\begin{tabular}{@{}c@{}} $\frac{R_\nu k^3}{54} -\frac{R_\nu k\omega^2}{48}$ \\ $ -\frac{3R_\nu R_\mathrm{b}^2 k\omega^2}{16(R_\nu-1)^2} $  \end{tabular} \\
        \hline
        $\delta_{\nueff}$  & $\frac{4 (1-R_\nu)}{3}$  & $\frac{2 R_\nu \omega}{3}$ & $\frac{2}{9} (R_\nu-1) k^2-\frac{R_\nu
   \omega^2}{4}$ & $\frac{R_\nu
   \omega^3}{12}-\frac{R_\nu k^2 \omega}{45} (R_\mathrm{b}+3)  $ \\
        \hline
        $\theta_{\nueff}/k$ & & $\frac{(1-R_\nu) k}{3}  $ & $\frac{1}{12} R_\nu k
   \omega$ & \begin{tabular}{@{}c@{}} $\frac{(R_\nu-1) (4 R_\nu+27) k^3}{54 (4
   R_\nu+15)} $ \\ $ -\frac{R_\nu k \omega^2}{48} $  \end{tabular} \\
        \hline
        $\sigma_{\nueff}$ & & & $-\frac{2 (R_\nu-1) k^2}{12
   R_\nu+45}$ &  $\frac{(61-4 R_\nu) R_\nu k^2 \omega}{18 (2
   R_\nu+15) (4 R_\nu+15)}$ \\
        \hline
        $F_{3{\nueff}}$ & & & & $-\frac{4 (R_\nu-1) k^3}{21 (4
   R_\nu+15)}$ \\
        \hline
%        $F_{4{\nueff}}$ & & & & \\
%        \hline
        $\delta_{\mathrm{b}}$ & $ -R_\nu$ & $\frac{R_\nu \omega}{2}$ & $\frac{1}{48}
   R_\nu \left(8 k^2-9 \omega^2\right)$ &\begin{tabular}{@{}c@{}} $ \frac{R_\nu \omega^3}{16} - \frac{1}{20 k^2\omega} $ \\ $-\frac{R_\nu R_\mathrm{b} (R_\nu-6)k^2\omega}{60 (R_\nu-1)} $  \end{tabular} \\
        \hline
        $\delta_{\DM}$ & $-R_\nu$ & $\frac{R_\nu \omega}{2}$ & $-\frac{3}{16}
   R_\nu \omega^2$ & $\frac{R_\nu
   \omega^3}{16}-\frac{R_\nu k^2 \omega (3 R_\mathrm{b}+4) }{180} $ \\
        \hline
        \end{tabular}
    \caption[Series evolution results for NID]{Series evolution for pure neutrino isocurvature $\varphi=\pi/2$.}
    \label{tab:series_evol_NID}
\end{table*}

\begin{table*}
    \centering
    \begin{tabular}{|c|c|c|c|c|c}
        \hline
        Perturbation & $\mathcal{O}(1)$ & $\mathcal{O}(\tau)$ & $\mathcal{O}(\tau^2)$ & $\mathcal{O}(\tau^3)$ \\
        \hline \hline
        $h$ & & $R_\DM\omega$ & $-\frac{3\omega^2}{8}R_\DM $ & $\frac{R_\DM}{360}(45\omega^3-16k^2\omega)$ \\
        \hline
        $\eta$ & & $-\frac{1}{6}R_\DM\omega$ & $\frac{\omega^2}{16}R_\DM$ & $-\frac{R_\DM}{240}(5\omega^3-4k^2\omega) $ \\
        \hline
        $\delta_{\gamma}$ &  & $-\frac{2}{3}R_\DM\omega$ & $\frac{\omega^2}{4}R_\DM$ & $-\frac{R_\DM}{60}(5\omega^3-4k^2\omega) $ \\
        \hline
        $\theta_{\gamma \mathrm{b}}/k$ & &  & $-\frac{1}{12}R_\DM k\omega$ & $\frac{R_\DM(3R_\mathrm{b}-R_\nu+1)k\omega^2}{48(1-R_\nu)} $
        \\
        \hline
        $\delta_{\nueff}$  &  & $-\frac{2}{3}R_\DM\omega$ & $\frac{1}{4}R_\DM\omega^2$ & $-\frac{R_\DM}{60}(5\omega^3-4k^2\omega)$ \\
        \hline
        $\theta_{\nueff}/k$ & &  & $-\frac{1}{12}R_\DM k\omega $ & $\frac{R_\DM}{48}k\omega^2$  \\
        \hline
        $\sigma_{\nueff}$ & & &  & \\
        \hline
        $F_{3\nueff}$ & & & &  \\
        \hline
%        $F_{4\nueff}$ & & & & \\
%        \hline
        $\delta_\mathrm{b}$& & $-\frac{1}{2}R_\DM\omega$ & $\frac{3\omega^2}{16}R_\DM$ & $-\frac{R_\DM}{80}(5\omega^3-4k^2\omega)$ \\
        \hline
        $\delta_\DM$ & $1$ & $-\frac{1}{2}R_\DM\omega$ & $\frac{3\omega^2}{16}R_\DM$ & $-\frac{R_\DM}{720}(45\omega^3-16k^2\omega)$ \\
        \hline
        \end{tabular}
    \caption{Series evolution for dark matter isocurvature $\varphi=0$.}
    \label{tab:series_evol_DMI}
\end{table*}

This is in contrast to the definition found in Ref.~\cite{Bucher:1999re}.
The initial conditions there are directly given in terms of gauge-dependent quantities and read
\begin{align}
    & \delta_{\nueff}(0)=1\,, \\
    & \delta_\gamma(0)=-\frac{\rho_{\nueff}}{\rho_\gamma}\,,  \\
    & \delta_\DM(0) =\delta_\mathrm{b}(0) =0\,.
\end{align}
Working backwards, this corresponds to
\begin{align}
    &S_{\gamma\nueff}=-\frac{3}{4}\left(1+\frac{\rho_{\nueff}}{\rho_\gamma}\right)\,, \\
    &S_{\gamma\DM}=S_{\gamma \mathrm{b}}=-\frac{3}{4}\frac{\rho_{\nueff}}{\rho_\gamma}\,.
\end{align}
To compare to the mixing angle formulation, we have to shift to $S_{\gamma\mathrm{b}}=0$ (cf. \cref{sec:overview neutrino isocurvature}). For this, the baryon and dark matter density perturbations need to be recast here as well. It follows that $\delta_\mathrm{b}=\frac{3}{4}\delta_\gamma$ 
and, with the compensated matter perturbation condition $\delta\rho_\mathrm{b}+\delta\rho_\DM=0$, 
$\delta_\DM=\frac{3}{4}\frac{\rho_{\nueff}}{\rho_\gamma}\frac{\rho_\mathrm{b}}{\rho_\DM}$. Retranslating into gauge-independent quantities yields
\begin{align}
    &{S}_{\gamma\nueff}
    =-\frac{3}{4}\left(1+\frac{\rho_{\nueff}}{\rho_\gamma}\right)\,,
    \\
    &{S}_{\gamma\DM}
    = -\frac{3}{4}\frac{\rho_{\nueff}}{\rho_\gamma}\left(1+\frac{\rho_{\mathrm{b}}}{\rho_\DM}\right) \,.
\end{align}
We refer to this setup as compensated neutrino isocurvature above, understanding it as a mixture of neutrino and DM isocurvature with the mixing angle
\begin{align}
    \varphi
    = \arctan\left(\frac{1+\frac{\rho_{\gamma}}{\rho_{\nueff}}}{1+\frac{\rho_{\mathrm{b}}}{\rho_\DM}}\right)\approx 1.12
    \,.
\end{align}
Note that also the normalization has to be adapted with a factor of
\begin{equation}
    \frac{3}{4}\frac{\rho_\nu}{\rho_\gamma}\sqrt{\left(1+\frac{\rho_\gamma}{\rho_\nu}\right)^2+\left(1+\frac{\rho_b}{\rho_\DM}\right)^2}\approx 1.4\,
    \label{eq:normalization}
\end{equation}
to compare to the parametrization proposed here.

The equations of motion feature diverging terms as one takes $\tau\rightarrow 0$. One therefore has to expand the solution to finite conformal time $\tau$ in order to solve them numerically. The two time scales of relevance are horizon entry at $\tau\approx k^{-1}$ as well as onset of matter domination at $\tau\approx\omega^{-1}$\,, where $\omega=\mathcal{H}_\mathrm{eq}/\sqrt{2}$ and $\mathcal{H}_\mathrm{eq}$ is the conformal Hubble rate at matter-radiation equality. The expansion for the perturbations for pure neutrino and pure dark matter isocurvature can be found in \cref{tab:series_evol_NID} and \cref{tab:series_evol_DMI}, respectively. For brevity of notation we introduce the following energy ratios
\begin{equation}
    R_{\nu}=\frac{\rho_{\nueff}}{\rho_\mathrm{rad}}\,,\quad R_{\DM/\mathrm{b}}=\frac{\rho_{\DM/\mathrm{b}}}{\rho_\mathrm{mat}}\,.
\end{equation}

%%%%%%%%%%%%%%%%%%%%%%%%%%%%%%%%%%%%%%%%%%%%%%%%%%%%
\section{LSS and CMB observables}
\label{sec:CMB polarization and LSS}
%%%%%%%%%%%%%%%%%%%%%%%%%%%%%%%%%%%%%%%%%%%%%%%%%%%%
\begin{figure}
    \centering
    \includegraphics[width=\columnwidth]{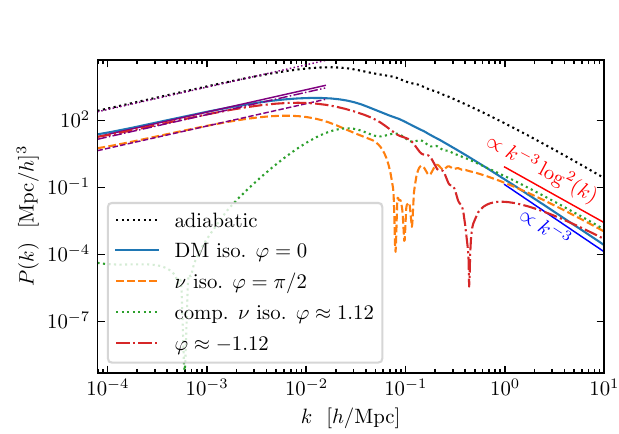}
    \caption{Comparison of the linear matter power spectrum today (redshift $z=0$) resulting from adiabatic, dark matter and neutrino isocurvature only as well as two admixed cases. For simplicity, we choose the scalar index $n=1$ and use the same amplitude $A_s$\,.}
    \label{fig:LSSspectra}
\end{figure}

For completeness we here present numerical results of the matter power spectrum and signatures of CMB polarization for the different modes of perturbations. We further give some analytic approximations applicable to the largest scales following Chs. 7+8 of Ref.~\cite{Dodelson:2003ft}. See also Ref.~\cite{Lyth:2009imm}.\\

\textbf{Large Scale Structure.}
It is well known that on scales that enter the horizon after matter-radiation equality, the curvature perturbation $\zeta_\mathrm{mat}$ corresponds to a constant Newtonian gravitational potential $\Phi$ and a matter perturbation $\delta_\mathrm{m}$ growing linearly in the scale factor as long as the contribution of dark energy to the total energy is negligible. 
When considering an adiabatic initial condition the curvature perturbation in matter domination is trivially related to the primordial one, $\zeta_\mathrm{mat}=\zeta_\mathrm{p}$\,.
In the case of vanishing primordial curvature perturbation but with isocurvature of the form
\begin{equation}
   S_{\gamma \DM}=S \cos{\varphi} \,,\quad S_{\gamma\nu}=S \sin(\varphi)\,,
\end{equation}
implying an isocurvature between the total radiation and matter 
\begin{equation}
    S_\mathrm{rad,mat}=S\left[R_\DM\cos(\varphi)-R_\nu\sin(\varphi)\right]\,,
\end{equation}
the curvature perturbation after matter-radiation equality can be found e.g. via the separate universe approach as \cite{Kodama:1986ud,Lyth:1998xn,Wands:2000dp}
\begin{equation}
    \zeta_\mathrm{mat}=-\frac{1}{3}S_\mathrm{rad,mat}\,.
\end{equation}
In general, the linear matter power spectrum resulting from scale invariant primordial fluctuations with amplitude $A_s$ for modes that enter the horizon long after matter-radiation equality may therefore be approximated as
\begin{equation}
    P(k,a)=A_s\left(\zeta_{\mathrm{mat},s} \frac{2k^2}{5(aH)^2}\frac{D_1(a)}{a\,\Omega_\mathrm{mat}}\right)^2\frac{2\pi^2}{k^3}\,,
\end{equation}
where the growth factor $D_1$ accounts for corrections to the linear growth with the scale factor $a$ due to dark energy. The factor $\zeta_{\mathrm{mat},s}$ captures the amount of curvature perturbation induced by a scalar mode $s$.
It is $\zeta_{\mathrm{mat},s=\zeta}~=~1$ in the adiabatic case and $\zeta_{\mathrm{mat},s=S}~=~-\left[R_\DM\cos(\varphi)-R_\nu\sin(\varphi)\right]/3$ for the generalized isocurvature mode.

In \cref{fig:LSSspectra} we show the linear matter power spectrum at redshift $z=0$ for the adiabatic and four different isocurvature modes resulting from flat primordial spectra with an amplitude $A_s=2.1\times 10^{-9}$ using the best fit values determined by Planck \cite{Planck:2018vyg} for the background cosmology. At large wavelengths we find excellent agreement with our analytic result (thin, purple lines). Note that for compensated neutrino isocurvature (green, dashed-dotted line), by definition $S_\mathrm{rad,mat}=0$ and therefore the leading contribution to the power spectrum proportional to $k$ is absent. 

A fully analytic discussion of the matter fluctuations on small scales that enter the horizon during radiation domination is beyond the scope of this work. We do note however that there are two solutions for the matter fluctuations during radiation domination. One is a constant and the second one is driven by gravitational interactions. Since fluctuations do not grow during radiation domination, the impact of these interactions quickly drops off after horizon entry and results in the second kind of solution growing logarithmically with the scale factor afterwards.

In the case of isocurvature fluctuations, the gravitational source terms, e.g. the Newtonian potential, vanish before horizon crossing. However, due to the difference in dynamics between neutrinos and the baryon-photon fluid, after horizon crossing, overdensities in the total radiation density build up and generate such terms. For the logarithmic growing solution to be present one therefore requires $S_{\gamma\nu}\neq 0$\,. Indeed we find that for the pure matter perturbation $\varphi=0$ the matter power spectrum falls proportional to $k^{-3}$ at large $k$, while for all other initial conditions eventually the logarithmically enhanced term proportional to $k^{-3}\log^2(k)$ comes to dominate as we have indicated in \cref{fig:LSSspectra}.

On the other hand, for compensated neutrino isocurvature, the constant solution seems to vanish. From our numerical studies we find that over the range $\varphi\in[\approx 1.12,\pi\sim 0]$ the power spectrum has one or multiple  zeros (presumably due to baryon acoustic oscillations). It seems reasonable that this is due to the two solutions interfering destructively with one of them vanishing and flipping sign at the limits of this interval.\\

\begin{figure}
    \centering
    \includegraphics[width=\columnwidth]{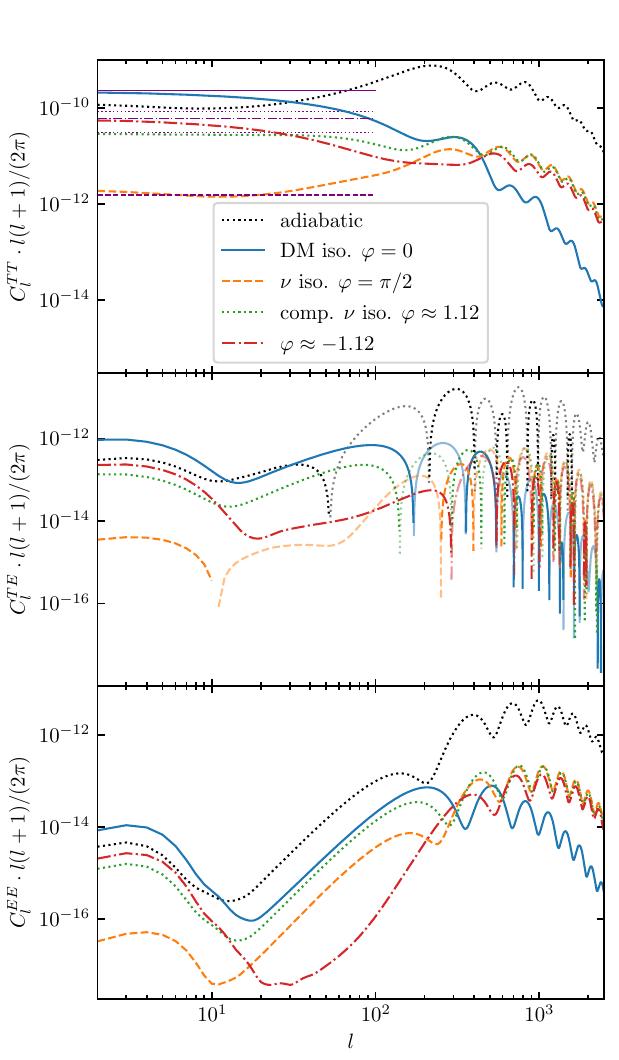}
    \caption{Same as \cref{fig:LSSspectra} but comparing correlations in CMB temperature and polarization. In the middle panel showing the cross-correlations between temperature and polarization, lighter colors indicate negative values.}
    \label{fig:CMBspectra}
\end{figure}

\textbf{CMB observables.}
As for the CMB observables we limit our analytic efforts to the Sachs-Wolfe effect describing the TT-correlations at small multipoles $l$ \cite{Sachs:1967er}.
One may show that in Newtonian gauge during matter domination, the temperature fluctuations $\Delta T/T=\delta_\gamma/4$ for super horizon modes are given by
\begin{equation}
    \frac{\Delta T}{T}=\frac{2}{3}\Phi+\frac{R_\DM}{3}S_{\gamma DM}\,.
\end{equation}
The Newtonian potential $\Phi$ is given by $\Phi=3/5\ \zeta_\mathrm{mat}$\,. After the photons start free-streaming, they additionally get redshifted as they climb out of the gravitational potential. The observed temperature fluctuation is therefore given by
\begin{align}
    \left(\frac{\Delta T}{T}\right)_\mathrm{obs}=\frac{\Delta T}{T}+\Psi&=-\frac{1}{5}\zeta_\mathrm{mat}+\frac{R_\DM}{3}S_{\gamma \DM}\\
    &=\frac{S}{15}\left[6R_\DM \cos(\varphi)-R_\nu \sin(\varphi)\right]\,,
\end{align}
where we used $\Psi\approx -\Phi$ for the second Newtonian potential.
The angular spectrum in the Sachs-Wolfe approximation is then given by
\begin{equation}
    \frac{C_{l}^{TT,\mathrm{sw}}\cdot l(l+1)}{2\pi}=\frac{A_s}{25}\left[\frac{6R_\DM}{3} \cos(\varphi)-\frac{R_\nu}{3} \sin(\varphi)\right]^2\,
\end{equation}
where in the case of adiabatic initial conditions one neglects the factor in square brackets.

In the top panel of \cref{fig:CMBspectra} we compare the analytic estimate (thin purple line) to the numerical results and find reasonable agreement for $l\lesssim 100$ given that we neglected all effects leading to a changing gravitational potential (integrated Sachs-Wolfe effect). For completeness we also show the spectra of CMB polarization.

\bibliography{apssamp}% Produces the bibliography via BibTeX.

\end{document}